\documentclass[12pt]{article}

\newtheorem{Theorem}{Theorem}[section]

\newtheorem{Proposition}[Theorem]{Proposition}

\usepackage[english]{babel}

\makeatletter
\usepackage{latexsym}
\usepackage[T1]{fontenc}
\usepackage{amsmath}
\usepackage{amssymb}

\def \be {\begin{equation}}
\def \ee {\end{equation}}

\def \ra {\rightarrow}

\def \eqq {\equiv}

\def \b {{\beta}}

\def \eps {{\varepsilon}}

\def \l {{\lambda}}

\def \A {{\cal A}}
\def \B {{\cal B}}
\def \C {{\cal C}}

\def \F {{\cal F}}
\def \G {{\cal G}}
\def \H {\mbox{${\cal H}$}}

\def \K {{\cal K}}

\def \O {{\cal O}}
\def \P {{\cal P}}

\def \T {{\cal T}}
\def \U {{\cal U}}

\def \Z {{\cal Z}}
\def \u {{\mathsf U}}

\def \abf {{\bf a}}

\def \x {{\bf x}}

\def \Rbf {{\bf R}}

\def \AO {{\cal A}({\cal O})}
\def \AO' {{\cal A}({\cal O}')}

\newfam\Ssfam
\font\eleSs=cmss10 at12pt \font\sevenSs= cmss10 at 8pt \font\sixSs= cmss10 at 6pt

\textfont\Ssfam=\eleSs\scriptfont\Ssfam=\sevenSs%
\scriptscriptfont\Ssfam=\sixSs \def\Ss{\fam\Ssfam\eleSs}

\def\doppio#1{{\rm I}\kern-.1667em{\rm #1}}

\def\Q{\text{Q}\kern-.52em
    \text{\vrule height1.5ex width.5pt depth0pt}\kern.45em}

\def \Zmath {{\mathchoice {\hbox{$\Ss\textstyle Z\kern-0.4em Z$}}
{\hbox{$\Ss\textstyle Z\kern-0.4em Z$}} {\hbox{$\Ss\scriptstyle Z\kern-0.25em
Z$}} {\hbox{$\Ss\scriptscriptstyle Z\kern-0.2em Z$}}}}

\def\Cmath{{\mathchoice{\hbox{$\rm\textstyle\text{\kern.35em\vrule
   height1.5ex width.5pt depth0pt\kern-.35em C}$}}
{\hbox{$\rm\textstyle\text{\kern.35em\vrule
   height1.5ex width.5pt depth0pt\kern-.35em C}$}}
{\hbox{$\rm\scriptstyle\text{\kern.35em\vrule
   height1.5ex width.3pt depth0pt\kern-.35em C}$}}
{\hbox{$\rm\scriptscriptstyle\text{\kern.35em\vrule
   height1.5ex width.2pt depth0pt\kern-.35em C}$}}}}

\def \be{\begin{equation} \displaystyle}
\def \ee{\end{equation}}

\def \A*{\mbox{$A^{*} $}}
\def \B*{\mbox{$B^{*} $}}
\def \C*{\mbox{$C^{*} $}}

\def \bea{\begin{eqnarray}}
\def \eea{\end{eqnarray}}

\def \b{\beta}

\def \l{\lambda}

\@addtoreset{equation}{section}

\def \be {\begin{equation} \displaystyle}

\def \ee {\end{equation}}

\def \ra {\rightarrow}
\def\AO {\mbox{${\cal A}({\cal O})$}}
\def\AO'{\mbox{${\cal A}({\cal O}')$}}
\def\O {\mbox{${\cal O}$}}

\def\A{\mbox{${\cal A}$}}

\def \ra{\rightarrow}

\def \eps {{\varepsilon}}

\def \O {{\cal O}}

\def \A {{\cal A}}
\def \AO {\A(\O)}
\def \AOl'{\A(\O_{loc}')}

\def \B {{\cal B}}

\def \F {{\cal F}}

\def \H {{\cal H}}

\def \P {{\cal P}}

\def \x {{\bf x}}

\def \FO  {\F_{obs}}

\begin{document}
\begin{titlepage}

\title{Why Gauges? Gauge symmetries for the classification of the  physical states*}

\sloppy

\author{F. Strocchi  \\  Dipartimento di Fisica, Università di Pisa, Pisa, Italy}

\fussy

\date{}

\maketitle

\begin{abstract}
This note focuses the problem of motivating the use of gauge symmetries (being  the identity on the observables) from general principles,  beyond their practical success, starting from global gauge symmetries and then by emphasizing the substantially different role  of local gauge symmetries. In the latter case, a deterministic time evolution of the local field algebra, necessary for field quantization,  requires a reduction of  the  full local gauge group $\G$  to a residual local  subgroup $\G_r$ satisfying suitable conditions.   A non-trivial residual local gauge group allows for a description/construction of the physical states by using  the vacuum representation of a local field algebra, otherwise precluded if $\G$ is reduced to the identity. Moreover, in the non-abelian case  the non-trivial topology of the a residual $\G_r$ defines the (gauge invariant) topological invariants which classify the vacuum structure with important physical effects; furthermore, it provides a general mechanism of spontaneous symmetry breaking without Goldstone bosons.

\end{abstract}

\vspace{5mm}



\vspace{5mm}

\thanks{*Tutorial article for EPJP}

\end{titlepage}

\section{Introduction}

The problem of motivating the use  of gauge symmetries from general principles, beyond  their practical success,  has drawn the attention of theoretical physicists and philosophers  of science and it is still a subject of investigations and debate. 

The introduction of (compact) \textit{global gauge groups} is motivated  by the description of states carrying superselected quantum numbers, which correspond to the spectrum of the invariant polynomial functions  of the  generators of a global  gauge group. A rigorous proof of this connection is provided by the deep results of Doplicher, Haag and Roberts \cite{Ha}, according to which, under general conditions,  localizable states carrying superselected quantum numbers are described by the vacuum representation  of a local field algebra carrying the representation of a compact Lie group. 

\textit{Local gauge groups} are asked for in connection with the description of states carrying superselected quantum charges related to local Gauss laws,  i.e. generated by currents which are the divergence of an antisymmetric tensor field, like the electric charge generated by the density of the  electric current, which satisfies  the Maxwell equations $j_\mu = \partial^\nu F_{\nu\, \mu}$. A general  argument in favor of local gauge symmetries in connection with Gauss' laws in presence of external charges  has been given in \cite{BCRV}  (see also  \cite{FS23} for a different motivation).

A delicate issue for the use of local gauge symmetries in quantum field theory is the need of reducing the symmetry under the full local gauge group $\G$, associated to a global gauge group $G$, in order to obtain  a \textit{deterministic time evolution} of the fields, which is required for their quantization. In fact, in quantum mechanics  the time evolution of quantum operators is necessarily deterministic, being described the  unitary operators $U(t) = \exp{(-i t H)}$. 

Thus,  in a quantum   field theory the symmetry under the full local gauge group $\G$ must be reduced to one of its local subgroups $\G_r$, which should not be the identity (as in the unitary gauge), since otherwise  the locality of the field algebra is lost (as in the Coulomb gauge of quantum electrodynamics \cite{FS16}), giving rise to basic technical problems. 

Conditions on the residual local group $\G_r$ for allowing  deterministic  field evolution are discussed  Section 3, subsection c).

Another important property of the  residual local group $\G_r$ in the non-abelian case is to provide  topological  invariants which classify the vacuum structure; the so emerging $\theta$ angle does not rely on the  instanton semiclassical approximation, but is rather given by the spectrum of the topological operators which describe the topology of $\G_r$. The derivation in Section 3, subsection d), without reference to gauge fixings,  generalizes that of the temporal gauge.

\section{The operational and the  mathematical \\ description of a physical system}

The operational/experimental description of a physical system is given by the  measurements of its observable quantities, and their time evolution,  on its physical states, prepared according to protocols of experimental preparations. 

The task of theoretical physics is to find a transcription of such an operational setting into a mathematical language and to provide models  and predictions.

Such an operational framework has an essentially unique mathematical transcription (see e.g. \cite{FS08, FS12}): the observables generate a $C^*-$ algebra $\A$ (with identitiy), the time evolution defines a one-parameter group of transformations (technically automorphisms) $\alpha_t$, $t \in \Rbf$, of $\A$, the experimental expectations of a   physical state $\omega$, $< A >_\omega$, $A \in \A$  define a 
positive linear functional on $\A$, $\omega(A)  = < A >_\omega$.  
  
By the Gelfand, Naimark, Segal theorem any state $\omega$ defines a representation $\pi_\omega$  of $\A$ in terms of (bounded) operators $\pi_\omega(A)$, $\forall A \in \A$, in a Hilbert space $\H_{\pi_\omega}$, with a cyclic vector $\Psi_\omega$ such that $\omega(A) = (\Psi_\omega,  \pi_\omega(A) \Psi_\omega)$ (with $ (\, , \, )$  the Hilbert scalar product in $H_{\pi_\omega}$). 

A concrete realization of such a framework is provided by  ordinary quantum mechanical systems: the algebra of observables is generated by the observable  canonical  positions and momenta (more precisely by their exponentials which generate  the Weyl $C^*$-algebra $\A_W$). By Stone-Von Neumann theorem  there is only one irreducible regular representation, apart from unitary equivalence, namely the standard Schroedinger  representation in $ L^2(\Rbf^N, dx)$, so that  all the (regular) physical states are described by square integrable functions or by mixed states defined in such a representation where  the position operators act as  multiplication operators and the momenta as space derivatives. 

Specific models are described by giving  the Hamiltonian $H$ as a self-adjoint  operator, defined as a  function of the observable  positions and momenta,  and the time evolution  is obtained by the corresponding one-parameter group $U(t) = \exp{(- i \,t\,H)} $ of unitary operators. 

More difficult is the concrete realization of the description of infinitely extended systems. 

The first problem is the characterization of the algebra of observables, because  in general one may not dispose of observable canonical variables for its generation, and in practice one resorts to non-observables fields/variables  for its construction.
\goodbreak

\newpage

Moreover, in general, it is not possible to define the time evolution of $\A$ in terms of  an Hamiltonian which does not involve auxiliary non-observable  operators and a constructive approach becomes a problem; the standard strategy is to define and derive the time evolution of the observables from that of the fields which generate $\A$.

Another qualifying condition is the local structure of the observables; since each observable  is identified by the experimental apparatus used for its measurement and physical operations and measurements are inevitably localized in bounded regions  of space, the algebra of observables 
is generated by algebras localized  in bounded space regions $V$. 

In the relativistic case, it is convenient to take as bounded regions in space time the  double cones $\O$ (\cite{Ha}, Chapter III). Thus, a basic role is played by the \textit{local algebras of observables}:   
 $$\A = \cup_V \A(V), \,\,\,\,\,\,\,\,\,\mbox{or}\,\,\,\,\,\,\,\A = \cup_\O \A(\O),$$
respectively. 

As emphasized by Haag \cite{Ha}, locality is a very important physical requirement for the description of infinitely extended systems, leading in particular  to the following algebraic conditions:
$$ [\, \A(V), \, \A(V')\,] = 0, \,\,\,\,\mbox{if}\,\,\,\,V \cap V' =\emptyset,$$
\be{[\, \A(\O), \, \A(\O')\,] = 0, \,\,\,\,\mbox{if}\,\,\O' \,\,\, \mbox{is spacelike w.r.t.}\,\,\O. }\ee

Another requirement, supported by the local structure, is that space translations, (respectively space time translations in the  relativistic case) 
define automorphisms of the local algebra, mapping the algebra localized in a bounded region into the algebra localized in the translated region (see e.g. \cite{FS19})

Finally, infinitely extended systems exhibit disjoint realizations or disjoint "phases" corresponding to states whose protocols of experimental preparations  are not related by physically realizable  operations (typical examples are the thermodynamical phases in the infinite volume limit or the disjoint representations corresponding to different values of symmetry breaking order parameters \cite{FS21}).

 Thus, the complete characterization of the physical representations of $\A$, equivalently of the set $\Sigma$ of all physical states, abstractly defined by morphisms of $\A$,  involve a  non-trivial construction.

 In conclusion, the mathematical description of a physical system is provided by the triple $(\A, \alpha_t, \Sigma)$, but for infinitely extended systems,  a constructive realization and control without the introduction non-observables variables/fields  is not at hand.

A class of important representations are those defined by pure homogeneous states characterized by the property of being  invariant under a subgroup of space translations and satisfying the cluster property. Prototypical  examples are the ground  or equilibrium states and most of the results of constructive  approaches have dealt 
 with such representations. In the case of relativistic systems the analysis and complete characterization of the vacuum representations of observable field algebras  have been worked out by Wightman (the so-called Wightman quantum field theory).

\section{Gauge symmetries and local field algebras for the description of  physical states}

The standard 
 (in particular the perturbative)  approach to the determination of the physical representations of $\A$, beyond the vacuum sector,  introduces  a  field algebra $\F$, generated by fields corresponding to  the particle states,   using the free theory as a guide. If such particle states carry a superselected quantum  number, the corresponding fields, which are supposed to obtain such states  from the vacuum, cannot be observable fields and therefore the so introduced field algebra is an extension of the observable field algebra $\FO$.

In this way, the difficult determination  of the representations of $\A $ beyond  the vacuum sector is reduced to the easier determination of the vacuum representation of a  field algebra $\F$. 

\vspace{2mm}

\noindent {\bf {\textit{ a)  Global gauge groups and  the description of physical states.}}}
\vspace{1mm}

The worked out examples of the above strategy show that the field algebra carries the representation of a global gauge group $G$ with the following role:
\vspace{1mm}

\noindent i) $G$ provides a simple \textit {characterization of the observable algebra} $\FO$  as  the subalgebra of $\F$ which is pointwise invariant under  $G$;  whenever possible,  especially in the relativistic case,  it is convenient to introduce  a field algebra satisfying locality, since this automatically guarantees the locality of its observable subalgebra; 
\vspace{1mm}
 
\noindent  ii) the vacuum representation of $\F$ contains  representations of $\FO$  defined by physical \textit {states carrying superselected quantum numbers} corresponding to the spectrum of the gauge invariant   functions of the elements of $G$;
\vspace{1mm}

\noindent iii) the representation of $G$ in $\F$ provides not only  the building blocks  for the representations  of $\A$, with the fields acting a intertwiners from the vacuum sector, but it also  allows to define the time evolution of the observables through the interaction between the fields.\footnote{This role of gauge symmetries of allowing the definition  of interactions has been emphasized by Rovelli \cite{R} as the motivation for the introduction of gauge symmetries. In our opinion, the description of the physical states, namely  the representations of the observables, looks a more fundamental and upstream motivation. }

The usefulness  of property i) is already displayed by the quantum mechanical description of a systems of $N$ identical particles. 

In fact, the particle positions and momenta are not observable operators and  the corresponding Weyl algebra $\A_W$ is not the observable algebra.
 The simple way of characterizing the algebra of observables $\A$ is to introduce  the   group of permutations  $\P$ and define $\A$  as the  subalgebra of $\A_W$ which is pointwise invariant under the  gauge group of permutations. 

Therefore, according to  ii),  the description of the possible states of identical particles is provided by  the irreducible representations of the algebra of observables corresponding to the representations of the permutation group; such representations are characterized  by the eigenvalues of the  characters, which are the invariant functions of the elements of  $\P$ (see \cite{D} and for  a general approach \cite{MS21}).

Property i) is particularly useful for the characterization of the  observable algebra  for an infinitely extended system, an otherwise non-trivial problem, which is solved by  introducing  a field algebra  and a global gauge group whose pointwise invariance property defines the observables.

A familiar example is provided by the free Bose  gas, described by (non-observable) canonical bosonic field operators  with the 
invariance under a  $U(1)$ global gauge group   selecting the observable (polynomial functions  of the  bosonic) fields (see e.g.\cite{FS19}).

The above role of global gauge groups for the description of the physical states has been proved to be a general result by Doplicher, Haag and Roberts, by exploiting locality (for a comprehensive account with the relevant references see \cite{Ha}); they considered the representations of the local observable algebra defined by localizable states,\footnote{A state $\omega_\O$ is localized in the double cone $\O$ if 
it behaves like the vacuum state $\omega_0$ on observables localized in the spacelike complement  $\O'$ of $\O$: namely 
$\omega_\O(A) = \omega_0(A), \,\,\,\,\forall A \in \A(\O') $.} carrying superselected quantum numbers, then, under general assumptions (see \cite{Ha}), there is a local field algebra which carries a representation of a \textit{compact gauge group} and  whose vacuum representation gives such representations.  

This explains why compact gauge groups play such an important in theoretical physics for the description  of particle states, in connection with their localizability.

\goodbreak

\vspace{3mm}

\noindent {\bf {\textit{ b)  Local  gauge groups and  the classification of physical states.}}}
\vspace{1mm}

Interesting physical states are characterized  by superselected quantum numbers or charges corresponding  to conserved currents $J_\mu$ which are the divergence of an antisymmetric tensor, $J_\mu = \partial^\nu G_{\nu\,\mu}, \,G_{\nu\,\mu} = - G_{\mu\, \nu}$. Since such charges    (called \textit{Gauss charges}), are linked to the Gauss flux of $G_{i \, 0}$ at space infinity,  
they are not  localizable and  the correspondent states  cannot be  described   by local states, i.e. by local fields applied to the  vacuum (see e.g. \cite{FS16}).
 
A prototypical example is provided by the electrically charged states of quantum electrodynamics (QED), the Gauss charge being the electric charge. Following Dirac's suggestion \cite{Dc},   such states    may  be   obtained by a non-local (infrared) dressing of the local states of the vacuum representation of a local field algebra\cite{BDMRS, FS16}.

As argued in \cite{FS23}, the use of a local field algebra $\F$ for the description/con\-struction of states which carry Gauss charges implies that $\F$ must carry a representation of a \textit{local gauge group}; this provides a physical motivation for the use of local gauge symmetries. 

Moreover, the local gauge group allows to identity/construct the local observable field algebra $\F_{obs}$ as the gauge-invariant subalgebra of  $\F$ (examples of  $\F_{obs}$ are the gauge-invariant polynomial functions of the fields which appear in the gauge invariant Lagrangian).

Given a  compact Lie gauge group $G$,  the corresponding local gauge group $\G$ is conventionally obtained by replacing the constant group parameters of $G$ by the set of infinitely differentiable functions, which is convenient to choose of compact support in space (for locality reasons)  and of at most polynomial increase in time, in order to guarantee their action as multipliers on the test functions of the smeared fields.  Such  regularity conditions are required for a mathematical well defined setting \cite{BS, FS16}.

\vspace{2mm}

\noindent {\bf {\textit{ c) Deterministic time evolution and invariance under a local gauge group  }}}
\vspace{1mm}

By a  general argument, not sufficiently emphasized in the literature, the local field algebra used for the construction of the physical states cannot carry a representation of the full local gauge group $\G$ correspondent to a global gauge group $G$.
 
For simplicity and concreteness,  we consider the Yang-Mills gauge theory, with the local gauge group $\G$ defined by $G$-valued $C^\infty$ unitary functions $\U$ differing from the identity only on a compact set in space, $\K_\U \subset \Rbf^3$; the corresponding gauge transformations are denoted by $\alpha_\U$. Hereafter, $G$ is assumed to be a simple Lie group.
\goodbreak 

Then, since  a dynamics defined by a $\G$ invariant  Lagrangian (or Hamiltonian) implies the $\G$ invariance of the equations of motion of the fields, the vector field $A(f))$ and its gauge transformed one $A^\U(f) \eqq  A(\U f \U^{-1}) + \U \partial \U^{-1}(f)$, satisfy  the same evolution equation for any $\U$.
 Moreover, one may choose $\U$  such  that its action reduces to the identity at the time $t_0$ which defines the initial condition of $A(f)$.
Hence,   we have two solutions of the same equations of motion, with the same initial conditions, so that \textit{determinism is lost}.

This implies that the local field algebra does not have a well defined time evolution and therefore  the fields cannot be defined as quantum operators, since in quantum mechanics the dynamics is defined by a unitary operator $U(t)$ and it is necessarily deterministic. 

By such an argument, one cannot   minimize the problem by invoking  the escape of a surviving deterministic evolution for the observable fields, as sometimes claimed in the literature; the whole strategy of using a local \textit{quantum} field algebra for obtaining the time evolution of its observable subalgebra and  the description/construction of physical states fails.

To overcome this difficulty one has to reduce the $\G$ invariance; the standard strategy is to add a gauge fixing term to a $\G$ invariant Lagrangian by still preserving the invariance under the global gauge group $G$ and under a local subgroup $\G_r \subset \G$; the motivation and advantages of such a choice shall be discussed below. A non-trivial residual local gauge group implies that the local Gauss laws  $J^a_\mu = \partial^\nu G^a_{\nu\, \mu}$ do not hold as operator equations in the vacuum representation of the local field algebra and one rather has
\be{J^a_\mu = \partial^\nu G^a_{\nu\, \mu} + G^a_\mu,}\ee
where $G^a_\mu$, called the \textit{Gauss operator}, depends on the choice of $\G_r$ and has vanishing expectations on the physical states,  which are   then  characterized by   Gauss charges.  

The extreme choice of reducing $\G$ to the identity implies the non-locality of the field algebra \cite{FS23}; a well known example is the Coulomb gauge quantization of electrodynamics.  The drawback is the difficult control of a non-local dynamics, the difficult construction of the local observable fields      as functions of non-local fields, involving  point splitting regularizations of products of non-local fields, the loss of the whole wisdom of local quantum field theory, including the renormalization procedure.

The natural question is how much   the local  gauge group $\G$,  parametrized by the  set of infinitely differentiable $G$-valued unitary functions $\U(\x, t)$ which differ from unity only on compact sets in space, should be reduced to avoid the loss of deterministic evolution of the field algebra.

\begin{Proposition} The condition of deterministic time evolution of the field algebra requires that the gauge functions $\U$ which define   the residual group $\G_r$, satisfy  a sort of deterministic  structure, namely
if two gauge functions   coincide at a given time $t_0$, together with their first order  time derivatives, then they coincide at any time $t.$ 
\end{Proposition}
\goodbreak
 In fact, by the above argument  $A(f)$ and $A^\U(f)$  satisfy the same initial conditions, if the action of  $\U$ reduces to the identity at a given time $t_0$, but then by the above condition $\U$ is the identity for any time, so that $A(f)$ and $A^\U(f)$ coincide at all times, with no violation of deterministic evolution. 

A residual local gauge group which leads to a deterministic time evolution of the  field algebra, will for brevity called a \textit{deterministic local gauge group}.

Examples of $V_r$ with such a deterministic structure  are 

\noindent 1) the $C^\infty$ functions $\eps$ which parametrize $\G_r$ are solutions of an hyperbolic equation, typically the free wave equation,  with initial data of compact support in space; this is the case of  the residual local gauge group $\G_r$ of the Feynman-Gupta-Bleuler (FGB) quantization of electrodynamics;

\noindent 2) the group parameters of $\G_r$ are solutions of the equation $\partial_t^2 \eps = 0$, with initial data of compact support in space, namely $\eps(\x, t) = c_1(\x)  + c_2(\x) \,t$.  

However, for a non-abelian residual group $\G_r$ the corresponding gauge parameters form an algebra with the product (denoted by $\wedge$) induced by $G$ invariance and  only  the case   with $c_2(\x) = 0$ is compatible with it  \cite{BS};
 this choice  characterizes the temporal gauge quantization of gauge field theories.  

For a non-abelian   $\G_r$  a condition yielding a consistent group  property 
is that its group parameters belong to the kernel of an operator $D$ having  the  property of a derivation (i.e. it satisfies the Leibniz rule), so that 
$$ D \eps_1 = 0, \,\,\,\,D \eps_2 = 0, \,\,\,\,\,\,\,\,\Rightarrow \,\,\,\,D (\eps_1 \wedge \eps_2) = (D \eps_1) \wedge \eps_2 + \eps_1 \wedge (D \eps_2) = 0.$$

\vspace{2mm}

\noindent {\bf {\textit{ d) Residual local gauge group and topological vacuum  structure}}}
\vspace{1mm}

As discussed above, the role of global gauge groups is to provide the superselected quantum numbers which classify the localizable physical states, and  local gauge groups are needed for a description/construction of physical states carrying a Gauss charge, through the vacuum representation of a local field algebra.

Moreover, a quantization with a residual (deterministic)  local  gauge group $\G_r$ allows for i)  the use of a local field algebra, ii) a simple identification  of the observable field subalgebra  by its  pointwise invariance under $\G_r$, iii) the exploitation of the Ward identities implied by $\G_r$, iv) the use of the local wisdom of renormalization,  v) a better control of the field dynamics, thanks to its locality.

In this subsection we shall argue that, in the non-abelian case, a residual tic) local gauge group  also provides the topological (gauge invariant) operators whose spectrum characterizes the topological vacuum structure. 

Such a very useful role is missed in quantizations with no residual local gauge group, like the BRST quantization or the unitary gauge quantization. In fact, the invariance group of the BRST quantization is not a (local) subgroup of $\G$ and it corresponds to infinitesimal transformations of the fields with parameters which are  no longer $C^\infty$ functions of compact support in space and of slow increase in time (i.e. multipliers of the test functions smearing the  singular operational character of the fields). Actually, the parameters of the BRST transformations are field operators and involve the problematic  multiplication of distributions at the same point, which is out of control beyond the perturbative approach; this precludes the possibility of considering  topological properties of the BRST transformations. Furthermore, the vacuum representations of the BRST field algebra does not satisfy positivity and in particular the ghost fields are not operators in a Hilbert space.      

\goodbreak
\def \bG  {\mathbb{G}}

\vspace{1mm}
\noindent \textit { $d_1$. Topology of gauge transformations and their  unitary local implementation.}

\vspace{1mm}

The following analysis does not make reference to specific gauge fixings and apparently generalizes the derivation of the results obtained in the temporal gauge \cite{FS19, FS21}.

Given the local gauge group $\G$ corresponding  to a (non-abelian) global gauge group $G$, (as defined above), we consider the local subgroup  of unitary gauge functions $\U(\x,t)$ which together  with  their first order derivatives have a limit $\u(\x) \in \G$,  for $|t| \ra \infty$ (possibly in some generalized sense). 

Actually, without much loss of generality, one might have included this property in the definition of the local gauge group $\G$,
 as the set of $G$-valued  $C^\infty$ unitary functions $\U(\x, t)$ which differ from unity only on a (time dependent) compact set $\K_{\U_t}  \subset \Rbf^3$ and have a $C^\infty$  limit, $\u \in \G$ together with its first derivatives, as $|t| \ra \infty$, with $\K_\u \eqq \K_{\U_\infty} \subset \Rbf ^3$    the compact set of its space localization; for simplicity, in the following, we shall adopt  this definition of $\G$ and denote by $\bG \subset \G$ the subgroup generated by such limits at infinite time.

The $C^\infty$ unitary gauge functions $\u(\x)$ differing from the identity  only on  compact sets $\subset \Rbf^3$ obviously extend to the one-point compactification $\dot{\Rbf}^3$  of $\Rbf^3$ and  define continuous mappings of $\dot{\Rbf}^3$ onto the global  gauge group $G$. 

Since $\dot{\Rbf}^3$ is isomorphic to $S^3$ and  each simple Lie group, in particular $SU(3)$, may be continuously deformed to one of its $SU(2)$ subgroups and $SU(2)$ is isomorphic to $S^3$, then, $\u(\x)$ defines a mapping of $S^3$ onto $S^3$. 

Such mappings  fall into disjoint homotopy classes labeled by the (topological invariant) {\bf \textit{winding number}} $n(\u)$  
\be{ n_\u = (24 \pi^2)^{-1} \int d^3 x \,\eps^{i j k} \,\mbox{Tr}\,[ \u_{(i)}(\x)\,\u_{(j)}(\x)\,\u_{(k)}(\x) ] \eqq \int d^3 x \,n_\u(\x),}\ee   
where  $\u_{(i)}(\x) \eqq \u(\x)^{-1} \partial_i\,\u(\x)$. 

\def \sfG   {\mathsf G}
In the following, $\u_n$ will denote a gauge function with winding number $n$; $ n=0$ characterizes the gauge transformations $\u$ which are contractible to the identity and the corresponding group  is denoted by $\bG_0$. A  gauge function $\U(\x, t) \in \G$ is  said to have winding number $n$ if so does its limit  $\u(\x)$.

Clearly, if a group parameter $\eps(\x, t)$ defines a gauge function $\U(\x, t)$ with zero winding number, so does also 
 $\lambda \eps(\x, t),  \lambda \in \Rbf$, and therefore $\eps(\x, t)$ determines a one-parameter group of unitary operators $V(\U_0(\lambda))$, for brevity sometimes denoted by $V(\U^\lambda)$.

Since $\bG_0$ is a normal subgroup of $\bG$, the quotient $\T = \bG/\bG_0$ is a well defined  abelian group, (with the product defined by the coset multiplication  $g\, \bG_0\,  h\, \bG_0 = g h \,\bG_0, \,g, h \in \bG$),  whose elements are the equivalence classes $\T_n$ labeled by the winding number $n$, with $\T_n \,\T_m = \T_{n+m}$.  
Then, $\T$ describes the  topological structure of $\bG$, and also of $\G$.  \goodbreak

The  relevance  of such a topological structure for the classification of the physical states shows up if the above  classification into homotopy classes applies also to the gauge functions of the (deterministic) residual gauge group $\G_r$, in particular  if $\bG$ is a subgroup of $\G_r$, as assumed for simplicity
 in the following.

  The strategy is somewhat similar to the standard approach to the vacuum structure in QCD, which classifies the topology of the instanton configurations in terms of their behavior at infinity, whereas now we make  reference  to  the well defined $C^\infty$ unitary gauge functions, rather than to  euclidean configurations with zero functional measure (see \cite{CO}, Appendix C; \cite {FS16}, Chapter 5, Section 8). 

The subgroup $\bG_0 \subset \G_r$, generated by gauge functions with $n = 0$, has a purely gauge content and it is reasonable to assume that the physical states (in particular the vacuum state) are invariant under it. Such a condition is indeed required for the physical interpretation of the temporal gauge.
  
 This implies that in a vacuum representation of the local field algebra $\F$  the gauge transformations  $\alpha_{\u_0},  \u_0 \in \bG_0$ are implemented by unitary operators $V(\u_0)$ (uniquely determined if the representation of $\F$ is irreducible). 

The localization property of the gauge functions, at the basis of our approach,  in turn implies that also the gauge transformations $\alpha_{\u_n}$ with $n \neq 0$ are implemented by unitary operators $V(\u_n)$. \goodbreak

In fact, the compact space support $\K_\u$ of the gauge functions $\u$ implies that $\alpha_\u (F) = F$, whenever the space support of the local field $F$ is disjoint from $\K_\u$. Therefore, if $\u_\abf(\x) \eqq \u(\x - \abf) $, for $\abf$ large enough, one has 
$$\omega_0(\alpha_\u(F))  =\omega_0( \alpha_\u \alpha_{\u^{-1}_\abf}(F)) = \omega_0(F),$$
since $\u \u^{-1}_\abf \in \bG_0$.   

The space localization of the gauge transformations implies that  the implementers $V(\u)$ commute with the   fields localized in double cones with space basis   disjoint from $\K_\u$.  
\vspace{1mm}

\noindent \textit{$ d_2.$ Topological group and  its representation in the center of the observables.}

\vspace{1mm}  
Guided by the previous discussion,  we consider a vacuum representation of a local field algebra $\F$ and its observable subalgebra in a Hilbert space $\H$; 
we further assume  that the physical state vectors are characterized by their invariance under the implementers  $V(\u_0)$ of the gauge transformations with zero winding number.  The projection  on the physical  subspace $\H' \subset \H$ is denoted by $P_0$.

This assumption is supported by the relation between $\bG_0 \subset \G_r$ and the Gauss operator, as  argued for  relevant gauge choices \cite{FS23}.

Then, we construct a representation of the topological group $\T$ in terms of unitary operator in the physical space.

To this purpose we introduce the following operators
\be { T_n \eqq P_0\,V(\u_n)\, P_0, \,\,\,\,\,\, T_n \,\H' \subset \H'.}\ee 
They satisfy the following properties \cite{MS09} \cite{FS21}:

\vspace{1mm}
\noindent 1) $T_n$ depends only on the equivalence class of $V(\u_n)$, since $V(\u'_n) = V(\u_n) V(\u_0)$, with $\u_0 = \u_n^{-1}\, \u_n' \in \bG_0$;

\vspace{1mm}
\noindent 2) $ T_n\, T_m = T_{n + m}$,

\noindent since $V(\u_0) V(\u) P_0 \Psi = V(\u) V(\u'_0) P_0 \Psi = V(\u) P_0 \Psi$, so that $V(\u_n) P_0  = P_0 V(\u_n) P_0 = ( P_0 V(\u_n)^*P_0)^* = (V(\u_n)^* P_0)^*  = P_0 V(\u_n)$, i.e.
$V(\u) \,P_0 =P_0\,V(\u)$; 
\vspace{1mm}

\noindent 3) $T_n\, T_n^* = T_n\,T_{-n} = T_0 = P_0$;
\vspace{1mm}

\noindent 4) $V(\u) T_n V(\u)^{-1} = T_n $ (gauge invariance);
\vspace{1mm}

\noindent 5) $T_n$ belongs to the local center $\Z$ of the observables (since it  commutes  with $P_0$). 

In conclusion,  the topological group $\T$ is represented by the gauge invariant operators $T_n$, which act as unitary operators on the physical subspace and define a non-trivial center of the local observables.  \goodbreak

\vspace{2mm}
\noindent \textit{ $d_3$. Topological vacuum structure and chiral symmetry breaking. \\ Another  mechanism of spontaneous symmetry breaking}
\vspace{1mm}

Since the factorial representations of the algebra of observables are characterized  by the points of the spectrum of its center,  they are labeled by the angle $\theta \in [0, \,\pi )$ which defines the spectrum of the abelian topological group of the
 form $\{ e^{ i 2 n \theta}, \theta \in [0, \pi) \}$. 

In this way, one obtains the topological vacuum structure (the so-called $\theta $ vacua) with a non-perturbative derivation.
The so derived vacuum structure is at the basis of important physical effects and provides  a prototype of  \textit{another  mechanism of spontaneous symmetry breaking beyond the Nambu-Goldstone mechanism and the Higgs mechanism}.

  In fact, quite generally if a continuous symmetry $\b^\l, \,\l \in \Rbf$ (technically an automorphism of the field algebra) does not leave the local center of the observables pointwise invariant, it is spontaneously broken in each irreducible/factorial representation of the observable algebra. 

Moreover, such a symmetry cannot be described by gauge invariant local  unitary implementers
$V_R(\l), \l \in \Rbf$, (localized in increasing spheres of radius $R$),  since then they would belong to the algebra of observables  and would  leave the center pointwise invariant.

Particularly interesting is the case in which the gauge transformations of the local implementers of $\b^\l$  are  given by phases: $ \alpha_\U(V_R(\l))  = V_R(\l)  e^{i v(\U, \l)}$; then the group law requires $v(\U, \l) = v(\U) \l$ and, if the vacuum state $\omega_0$ is invariant under the gauge transformations $\alpha_\U$, the unitary implementers $V_R(\l)$ cannot be weakly continuous in $\l$. In fact one has, $ \forall A \in \A$ 
\be{ \omega_0(V_R(\l) A) = \omega_0( \alpha_\U(V_R(\l) A)) = \omega_0(V_R(\l)A) e^{i v (\U)\,\l},}\ee
which implies $\omega_0(V_R(\l) A) = 0,$ for $\l \neq 0$, whereas $\omega_0(V_R(\l) A) = \omega_0( A ), $ for $\lambda = 0$.

\begin{Proposition} Under the above conditions, the spontaneous breaking of the symmetry $\b^\l$ does not requires the existence of associated Goldstone bosons.
\end{Proposition}  

In fact, the basic assumption of the Goldstone theorem is that $\b^\l$ is locally generated by a conserved local current, in the sense that on the vacuum the infinitesimal transformation $\delta A$ of the operator $A$, which gives the symmetry breaking order parameter, is given, for $R$ large enough, by 
$$ < \delta A > = \frac{d}{ d\l} < \b^\l(A) >|_{\l =0} = i <  [\, J_0(f_R, \alpha), \,A\,] > = < \frac{d}{ d \l } V_R(\l)\,  A - A\, \frac{d}{d \l} V_R(\l) >  $$
\be{ =  2 i\, \mbox{Im} <  \frac{d}{d \l}  V_R(\l)  A >,}\ee
 the Goldstone spectrum being given by the expectation on the right hand side.\cite{FS21}  

The validity of such an equation is precluded by the lack of weak continuity in $\l$ of $V_R(\l)$, which excludes  the existence of the derivative with respect to $\l$. Hence the Goldstone theorem does not apply.

Such a mechanism applies to the breaking of chiral symmetry in QCD, $\b^\l$, $\l \in \Rbf$, which may be shown to be locally implemented by unitary operators $V^5_R(\l)$, formally the exponentials $\exp( i \l \,J_0^5(f_R, \alpha))$    of the density $J_0^5$ of the conserved gauge dependent chiral current $J_\mu^5 = j^5_\mu + 2 C_\mu, \,\,\,$, where $j^5_\mu $ is the gauge invariant chiral current with anomaly and $C_\mu$ is the topological current \cite{FS19, FS21}. 

The quantum version of  the classical gauge transformations 
$$\alpha_\u (C_0) = C_0 -(8\,\pi^2)^{-1} \partial_i \eps^{i j k} \mbox{Tr} [  \partial_j \u(\x) \, \u(\x)^{-1}  A_k]$$ 
gives 
\be{\alpha_{\u_n}(V^5_R(\l)) = V(\u_n) V^5_R(\l) V(\u_n)^{-1}  = 
e^{i 2 n  \l } \,V^5_R(\l)}\ee
for any $f_R$ such that $f_R = 1 , $ on $\K_\u$ (so that  $f_R \partial_j \u = \partial_j \u,  \partial_i \,f_R \partial_j \u = 0$). 
This implies that $V_R^5(\l)$ is not weakly continuous in $\l$.  

Furthermore, thanks to the localization properties of the $V(\u_n)$, the  chiral transformations of the $V(\u_n) $ are given by the action of the local implementers yielding (by eq.\,(3.6))
\be{ \b^\l( V(\u_n)) = e^{- 2i n \l}  \,V(\u_n).}\ee  
This leads to $\b^\l(T_n) = e^{ -2i n \l}\,T_n$ (for a detailed argument see \cite{MS09}), i.e. the center is not pointwise invariant under the chiral symmetry. Then, Proposition 3.2 applies  and one gets a solution of the $U(1)$ problem.

\vspace{10mm}


 \end{document}